\documentclass[12pt,fleqn,a4paper]{article}

\usepackage{a4wide}
\usepackage{epsfig}
\usepackage{graphicx}
\usepackage{multicol}
\usepackage{slashed}
\usepackage[intlimits]{amsmath}
\usepackage{amssymb}
\usepackage[small]{caption}
\usepackage[usenames,dvipsnames]{color}
     \definecolor{hgreen}{rgb}{0,.3,0}
     \definecolor{hred}{rgb}{.3,0,0}
     \definecolor{hblue}{rgb}{0,0,.3}
     \definecolor{LightGray}{gray}{0.95}
\usepackage[
            colorlinks=true,
            linkcolor=hblue,
            citecolor=hgreen,
            filecolor=hblue,
            urlcolor=hred]{hyperref}

\numberwithin{equation}{section}

\providecommand{\doi}[1]{\href{http://dx.doi.org/#1}{doi:#1}}

\newcommand{\MS}{\overline{\text{MS}}}

\newcommand{\MZ}{M_{Z}}

\newcommand{\xt}{x_{t}}

\newcommand{\muc}{\mu_{c}}

\begin{document}


\title{
  \boldmath
    \textbf{Two-Loop Electroweak Corrections\\ for the
	    $K\to\pi\nu\bar\nu$ Decays}
  \unboldmath
}

\author{
{$\text{Joachim Brod}^{a,b}$, $\text{Martin Gorbahn}^{a,b}$, and $\text{Emmanuel Stamou}^{a,b,c}$}\\[2em]
  {\normalsize $^{a}$Excellence Cluster Universe, Technische Universit\"at M\"unchen,}\\
  {\normalsize Boltzmannstra\ss{}e 2, D-85748 Garching, Germany}\\[1em] 
  {\normalsize $^{b}$Institute for Advanced Study, Technische Universit\"at M\"unchen, }  \\ 
  {\normalsize  Lichtenbergstra\ss{}e 2a, D-85748 Garching, Germany}\\[1em] 
  {\normalsize $^{c}$Physik-Department, Technische Universit\"at M\"unchen,}\\
  {\normalsize James-Franck-Stra\ss{}e, D-85748 Garching, Germany} 
}
\date{September 2010}

\maketitle

\begin{abstract}
\addcontentsline{toc}{section}{Abstract}

The rare $K\to\pi\nu\bar\nu$ decays play a central role in testing the
Standard Model and its extensions. Upcoming experiments plan to
measure the decay rates with high accuracy. Yet, unknown higher-order
electroweak corrections result in a sizeable theory error. We remove
this uncertainty by computing the full two-loop electroweak
corrections to the top-quark contribution $X_t$ to the rare decays
$K_L\to\pi^0\nu\bar\nu$, $K^+\to\pi^+\nu\bar\nu$, and $B\to
X_{d,s}\nu\bar\nu$ in the Standard Model. The remaining theoretical
uncertainty related to electroweak effects is now far below
1\%. Finally we update the branching ratios to find
$\text{Br}(K_L\to\pi^0\nu\bar\nu) = 2.43(39)(6) \times 10^{-11}$ and
$\text{Br}(K^+\to\pi^+\nu\bar\nu) = 7.81(75)(29) \times 10^{-11}$. The
first error summarises the parametric, the second the remaining
theoretical uncertainties.

\end{abstract}


\section{Introduction}\label{sec:introduction}
    
The branching ratios of the rare $K^+ \to \pi^+ \nu \bar{\nu}$ and
$K_L \to \pi^0 \nu \bar{\nu}$ decays are dominated by contributions of
internal top-quarks in the Standard Model. This short distance
sensitivity results in a precise theory prediction, but also in a
proportionality to powers of $V_{ts}^* V_{td}$. Accordingly, the
branching ratios are suppressed with respect to generic new physics
scenarios by the near diagonality of the Cabibbo-Kobayashi-Maskawa
(CKM) matrix. This leads to a high sensitivity to new physics, and a
precision measurement of these modes could provide a decisive test of
the Standard Model and its extensions.

This potential will be exploited by a new generation of experiments
(NA62 at CERN, KOTO at JPARC, and the proposed future experiment P996
at Fermilab), which aim at measuring the branching ratios with
unprecedented precision.

In the Standard Model the $K \to \pi \nu \bar\nu$ decays proceed
through $Z$-penguin and electroweak box diagrams
which exhibit a power-like GIM mechanism. This implies a suppression
of non-perturbative effects and, related to this, that the low-energy
effective Hamiltonian \cite{Buchalla:1993wq,Buchalla:1998ba}
\begin{align}
\label{eq:HeffSM}
\mathcal{H}_{\text{eff}} = \frac{4G_F}{\sqrt{2}}
\frac{\alpha}{2\pi\sin^2\theta_{\text{W}}} \sum_{l=e,\mu ,\tau} \left(
\lambda_c X^l 
 + \lambda_t X_t \right) (\bar{s}_L \gamma_{\mu}
d_L) (\bar{\nu_l}_L \gamma^{\mu} {\nu_l}_L) + \text{h.c.}
\end{align}
involves to an excellent approximation only a single effective
operator. Here $G_F$ is the Fermi constant, $\alpha$ the
electromagnetic coupling and $\theta_{\text{W}}$ the weak mixing
angle. The sum is over all lepton flavours, $\lambda_i=V^*_{is}V_{id}$
comprise the CKM factors, and $f_L$ represents left-handed fermion
fields.

The functions $X^l$ constitute the charm-quark contribution to
$\mathcal{H}_{\text{eff}}$ and add 30\% to the total branching ratio
of the $K^+ \to \pi^+ \nu \bar \nu$ decay, while they leave the CP
violating $K_L \to \pi^0 \nu \bar \nu$ decay unaffected. The
theoretical uncertainty in $X^l$ is $2.5\%$ after
next-to-next-to-leading order (NNLO) QCD \cite{Gorbahn:2004my,
  Buras:2005gr,Buras:2006gb} and next-to-leading order (NLO)
electroweak corrections \cite{Brod:2008ss} are taken into account, and
the resulting error in the branching ratio is small.

The situation is different for the function $X_t$ which includes
internal top-quark loops: it gives either the sole or the dominant
contribution to the neutral or the charged decay modes,
respectively. A two-loop electroweak calculation should cancel the
sizeable scheme dependence of the input parameters. Yet, only NLO QCD
corrections \cite{Buchalla:1998ba, Buchalla:1992zm, Misiak:1999yg} and
the leading term of the large-$m_t$ expansion of the two-loop
electroweak corrections are known. While unknown higher-order QCD
corrections result in a 1\% uncertainty in $X_t$, the uncertainty
related to unknown sub-leading electroweak contributions is estimated
to be $\pm 2\%$~\cite{Buchalla:1997kz}. This can be understood in the
following way: the matching calculation with internal top-quark loops
is purely short distance, the resulting operator renormalises like a
current, such that the QCD perturbation theory converges well. Yet the
on-shell scheme counterterm of $\sin \theta_\textrm{W}$ includes large
higher terms in the large-$m_t$ expansion. Hence the renormalisation
scheme dependence of $\alpha/\sin^2\theta_{\text{W}}$ in
(\ref{eq:HeffSM}) cannot cancel if only the leading term in the
large-$m_t$ expansion is taken into account. This was found in
Reference~\cite{Buchalla:1997kz} where the scheme difference between the
on-shell scheme and the $\overline{\textrm{MS}}$ scheme was only
decreased from 5.6\% to 3.4\% through the inclusion of the first order
in the large-$m_t$ expansion.

In this paper we will improve on the analysis of
Reference~\cite{Buchalla:1997kz} and compute the full electroweak two-loop
corrections to the top-quark contribution $X_t$. Only in such a way is
it possible to fix the definition of the electroweak input parameters
and reduce the uncertainty due to unknown higher order electroweak
corrections from 2\% to the per mil level. Since a 2\% uncertainty in
$X_t$ scales up to a 3\% to 4\% uncertainty in the branching ratios
such a reduction of the theoretical error is important in particular
in light of the coming experiments. In addition, our results are
equally applicable for the $B\to X_{d,s}\nu\bar\nu$ decays.

Our paper is organised as follows. In Section~\ref{sec:structure} we
discuss the dependence of our result on different renormalisation
schemes. In Section~\ref{sec:calculation} we present some technical
details of our calculation. Our numerical results are contained in
Section~\ref{sec:numerics}. In the Appendices we provide the analytic
form of the electroweak correction to $X_t$ in the limit of small
$\sin\theta_W$ and compare our expansion for a large top-quark mass
with the literature.


\boldmath
\section{\texorpdfstring{$X_t$ beyond leading order}
			{X_t beyond leading order}}\label{sec:structure}
\unboldmath

The truncation of the perturbation theory results in a residual scale
and scheme dependence of the matrix elements of the effective
Hamiltonian in Equation~\eqref{eq:HeffSM}. For the top-quark contribution,
the matrix element of the operator
\begin{equation}
  \label{eq:Qnu}
  Q_{\nu} = \sum_{l=e,\mu,\tau}
  (\bar{s}_L \gamma_{\mu} d_L) (\bar{\nu_l}_L \gamma^{\mu} {\nu_l}_L)  
\end{equation}
factorises and $4\alpha G_F/(2\sqrt{2}\pi \sin^2\theta_W) \lambda_t
X_t$ will be independent of the renormalisation procedure after
higher-order corrections are included. Let us now discuss the
dependence on the electroweak renormalisation scheme and how to
combine these schemes with the NLO QCD results, which are known in the
$\overline{\textrm{MS}}$ scheme.

Pure QCD corrections leave $G_F$, $\alpha$, and $\sin^2\theta_W$
unaffected, such that $X_t$ is a renormalisation scheme invariant
quantity if electroweak effects are ignored. It is then customary to
expand
\begin{equation}
  X_t = X_t^{(0)} + \frac{\alpha_s}{4\pi}X_t^{(1)} +
  \frac{\alpha}{4\pi}X_t^{(EW)}
  \label{eq:calXfunction} \, .
\end{equation}
in terms of the leading-order (LO) contribution~\cite{Inami:1980fz}
\begin{equation}
  X_t^{(0)}=\frac{\xt}{8}\left[
    \frac{\xt+2}{\xt-1}+\frac{3\xt-6}{(\xt-1)^2} \ln \xt \right] \, ,
  \label{eq:calX0}
\end{equation}
where $x_t=m_t^2/M_W^2$. The schemes for $m_t$ and $M_W$ are defined
below. The NLO QCD
correction~\cite{Buchalla:1998ba,Buchalla:1992zm,Misiak:1999yg}
\begin{equation}
\begin{split}
  X_t^{(1)} & = -\frac{29 x_t - x_t^2 - 4 x_t^3}{3 (1 - x_t)^2} -
  \frac{x_t
    + 9 x_t^2 - x_t^3 - x_t^4}{(1 - x_t)^3} \ln x_t \\
  & + \frac{8 x_t + 4 x_t^2 + x_t^3 - x_t^4}{2 (1 - x_t)^3} \ln^2 x_t
  - \frac{4 x_t - x_t^3}{(1 - x_t)^2} \text{Li}_2 (1 - x_t) + 8 x_t
  \frac{\partial X_t^{(0)}}{\partial x_t} \ln \frac{\mu_t^2}{M_W^2} \,
  ,
\end{split}
\end{equation}
fixes the renormalisation scheme of the parameters which appear in the
LO contribution: namely, the top-quark mass. Here, the QCD part of the
top-quark mass counterterm is defined in the $\overline{\textrm{MS}}$
scheme.

The leading term in the large-$m_t$ expansion of the two-loop
electroweak corrections $X_t^{(EW)}$ can be found in
Reference~\cite{Buchalla:1997kz}, while the hitherto unknown full two-loop
result is computed in this paper. The sum of $X_t^{(0)}$ and
$X_t^{(EW)}$ will only be invariant under an electroweak scheme change
if it is multiplied by the normalisation factor of the effective
Hamiltonian, $4\alpha G_F/(2\sqrt{2}\pi \sin^2\theta_W)$.
Accordingly, the electroweak renormalisation scheme has to be fixed
for the parameters in the normalisation factor.

Since in the electroweak theory not all parameters are independent, we
have to specify the physical input parameters, and we choose the set
\begin{equation}\label{eq:OSinput} G_F, \,
  \alpha, \, M_Z, \, M_t, \, \text{and} \, M_H .
\end{equation}
Here $G_F$ is the experimental value of the Wilson coefficient
relevant for muon decay, $\alpha$ the fine structure constant, and
$M_Z$ the $Z$-boson pole mass.  $M_t$ is the top-quark mass, where QCD
corrections are renormalised in the $\overline{\textrm{MS}}$ scheme,
while the on-shell scheme is used for the electroweak corrections. The
Higgs mass $M_H$ is essentially a free parameter -- its value is
assumed to be consistent with electroweak precision data.

For fixed input parameters we can now study the remaining residual
higher-order uncertainty by using different renormalisation
schemes. In the following discussion we will make use of three
renormalisation schemes:
\begin{itemize}
\item The $\overline{\textrm{MS}}$ scheme for all parameters,
\item the on-shell scheme for all masses and the
  $\overline{\textrm{MS}}$ scheme for all coupling constants,
\item or the on-shell scheme for all masses and the weak mixing angle
  -- the QED coupling constant is renormalised in the
  $\overline{\textrm{MS}}$ scheme.
\end{itemize}
The explicit result for $X_t^{(EW)}$ is different for each
renormalisation scheme. In practise, we perform our calculation in the
$\MS$ scheme and transform our result into the respective scheme by a
finite renormalisation.

In all three schemes we renormalise the CKM parameters in the $\MS$
scheme and use $G_F$ as a normalisation factor for the effective
Hamiltonian in Equation~\eqref{eq:HeffSM}. The parameter $G_F$ plays a
special role, because it is by itself defined as a Wilson coefficient,
of the operator $Q_{\mu} = (\bar{\nu_\mu}_L \gamma_{\rho}\mu_L) ({\bar
  e}_L \gamma^{\rho} {\nu_{e}}_L)$ which induces the muon decay in the
effective Fermi theory. To make this more explicit we introduce the
following notation: We denote the Wilson coefficient for muon decay by
$G_{\mu} = G_{\mu}^{(0)} + G_{\mu}^{(EW)} + \ldots$, where the
superscript $(0)$ denotes the tree level contribution, $(EW)$ the
one-loop electroweak corrections, and the ellipses stand for terms
beyond second order in the electroweak interactions. By $G_F$ we then
denote the experimental value of $G_{\mu}$ as extracted from muon
life-time experiments~\cite{Marciano:1988vm,vanRitbergen:1999fi}. If
we now write the effective Hamiltonian~\eqref{eq:HeffSM} in the
general form
\begin{equation}
  \mathcal{H}_{\text{eff}} = \frac{4}{\sqrt{2}}
  \frac{\alpha}{2\pi\sin^2\theta_{\text{W}}} C_\nu Q_\nu =
  \frac{4G_F}{\sqrt{2}} 
  \frac{\alpha}{2\pi\sin^2\theta_{\text{W}}} X_t Q_\nu \, ,
\end{equation}
we find
\begin{equation}
X_t^{(0)} = \frac{C^{(0)}}{G_{\mu}^{(0)}} \, , \quad 
X_t^{(EW)} = \frac{C^{(EW)}}{G_{\mu}^{(0)}} -
  \frac{C^{(0)}G_{\mu}^{(EW)}}{\big(G_{\mu}^{(0)}\big)^2} \, . 
\end{equation}

\boldmath
\subsubsection*{The $\MS$ Scheme}
\unboldmath

In the $\MS$ scheme we use
\begin{equation}\label{eq:MSinput}
g_1, \, g_2, \, v, \, \lambda, \, \text{and} \, y_t
\end{equation}
as fundamental parameters. Here $g_1$ and $g_2$ are the couplings of
the $SU(2)$ and $U(1)$ gauge group, respectively, $v$ is the vacuum
expectation value of the Higgs field, $\lambda$ the quartic Higgs self
coupling, and $y_t$ the Yukawa coupling of the top quark. All these
parameters are running parameters, depending on the renormalisation
scale $\mu$. We fix the initial conditions of these parameters by
expressing the physical parameter set~\eqref{eq:OSinput}
through~\eqref{eq:MSinput} using one-loop accuracy\footnote{For the
  Higgs boson mass we use the tree-level relation.} and fitting the
values of~\eqref{eq:MSinput} to yield the experimental values
of~\eqref{eq:OSinput}.

We choose to cancel all tadpole diagrams with a finite
counterterm. This results in an additional finite renormalisation of
all massive quantities -- a sample diagram is shown in
Figure~\ref{fig:twoloop}. In this way we ensure that intermediate
results are gauge parameter independent.

\boldmath
\subsubsection*{Masses in the On-Shell Scheme}
\unboldmath

As a more well-behaved alternative, we use the on-shell definition of
the $W$-boson and the top-quark mass. Since we performed our
calculation in the $\MS$ scheme, we have to perform a finite mass
renormalisation. The necessary renormalisation constants consistent
with our treatment of tadpole diagrams can be found
in~\cite{Jegerlehner:2001fb,Jegerlehner:2002em}. 

In addition, we have to specify the renormalisation scheme for the
weak mixing angle. We will use the following two schemes:
\begin{itemize}
\item In the on-shell scheme the weak mixing angle is defined by
  $s_W^2 \equiv \sin^2\theta_W^{\text{on-shell}} = 1 -
  M_W^2/M_Z^2$. Here the $W$-boson mass is calculated including
  radiative corrections from the input parameter
  set~\eqref{eq:OSinput}, which introduces a Higgs-mass dependence. In
  addition, the use of the on-shell value for $\sin^2\theta_W$ implies
  a finite renormalisation of our $\MS$ results by including a finite
  counterterm for $\sin^2\theta_W$. It is given in terms of the
  on-shell renormalisation constants for $M_W$ and $M_Z$ by
  \begin{equation}
    \delta s_W = \frac{c_W^2}{2s_W} \left( \frac{\delta M_Z^2}{M_Z^2} -
      \frac{\delta M_W^2}{M_W^2} \right) \bigg|_{\Delta = 0} \, ,
  \end{equation}
  where the subscript $\Delta = 0$ implies setting the pole part
  including the finite subtraction, $\Delta = 1/\epsilon - \gamma_E +
  \log 4\pi$, to zero. The expressions for $\delta M_Z^2$ and $\delta
  M_W^2$ can again be found
  in~\cite{Jegerlehner:2001fb,Jegerlehner:2002em}.
\item The $\MS$ definition of the weak mixing angle, denoted by $\hat
  s_\text{ND}$, leads to numerically tiny NLO corrections. It is given in
  terms of $s_W^2$ by~\cite{Degrassi:1996ps}
  \begin{equation}
    \hat s^2_\text{ND} \equiv \sin^2\theta_W^{\MS} = s_W^2 \left( 1 +
      \frac{c_W^2}{s_W^2} 
      \frac{4\pi\hat\alpha(M_Z)}{\hat s^2_\text{ND}} \Delta\hat\rho
    \right)  \, ,
  \end{equation}
  where $\hat\alpha = \alpha^{\MS}$, and $c_W^2 = 1 - s_W^2$.  The
  explicit expression for $\Delta\hat\rho$ can also be found
  in~\cite{Degrassi:1996ps}.
\end{itemize}
The numerical discussion of the three different schemes is given in
Section~\ref{sec:numerics}.

\section{Calculation}\label{sec:calculation}

\begin{figure}[t]
  \begin{center}
    \includegraphics[width=0.8\textwidth]{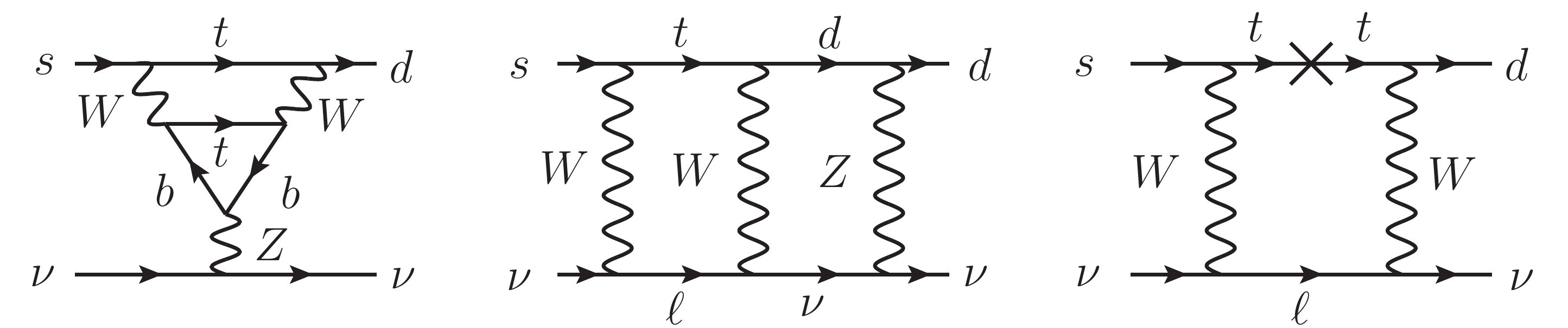}
      \caption{Sample penguin, box, and counterterm diagrams. Our
        tadpole renormalisation results in an explicit finite
        renormalisation of all massive quantities. The 
        right-hand side diagram shows a resulting 
        counterterm diagram. 
        \label{fig:twoloop}}
  \end{center}
\end{figure}

We determine the effective Hamiltonian by computing the relevant
Standard Model Green's functions in the $\MS$ scheme and matching them
to the five-flavour effective theory. To this end we have to calculate
two-loop box and penguin diagrams, samples of which are shown in
Figure~\ref{fig:twoloop}.
All diagrams reduce to two-loop vacuum diagrams after setting external
momenta and light masses to zero. The resulting loop integrals are
computed using standard methods
\cite{Bobeth:1999mk,Davydychev:1992mt}. All this is done in two
independent setups: one is using the FeynArts \cite{Hahn:2000kx}
package to generate the diagrams and a self written Mathematica
program, the other method uses a self written Form
\cite{Vermaseren:2000nd} program. The Feynman gauge $\xi=1$ is used in
both setups.

The integrals in the effective theory correspond to massless diagrams
with vanishing external momenta and are exactly zero in dimensional
regularisation. The only remaining contributions are then products of
renormalisation constants and tree-level matrix elements of the
operators $Q_\nu$, defined in Equation~\eqref{eq:Qnu}, and
\begin{equation}\begin{split}\label{eq:Qs}
    E_{\nu} & = \sum_{l=e,\mu,\tau} (\bar{s}_L
    \gamma_{\mu_1}\gamma_{\mu_2}\gamma_{\mu_3} d_L) 
    (\bar{\nu_l}_{L}
    \gamma^{\mu_1}\gamma^{\mu_2}\gamma^{\mu_3} {\nu_l}_{L}) -
    (16-4\epsilon) Q_{\nu}\, .
\end{split}\end{equation}
The evanescent operator $E_{\nu}$ arises in the context of dimensional
regularisation and vanishes algebraically in four space-time
dimensions. It leads to a non-vanishing finite contribution to the
Wilson coefficient, proportional to the finite mixing of $E_{\nu}$
into $Q_{\nu}$. The infinite operator renormalisation constants are
determined from the ultraviolet poles of the matrix elements of the
operators between external fermion states. They multiply the
tree-level and one-loop Wilson coefficients of the
operators~\eqref{eq:Qnu}~and~\eqref{eq:Qs} and cancel exactly the
corresponding spurious infrared divergences of the Standard Model
amplitude, thus rendering the matching condition finite.

The use of dimensional regularisation is in general inconsistent with
a fully anticommuting $\gamma_5$ matrix in $d$ dimensions, and we use
the 't Hooft-Veltman (HV) scheme in our calculation. However, problems
only arise when computing traces containing at least three $\gamma_5$
matrices, appearing in the anomalous fermion triangles (see for
instance the first diagram in Figure~\ref{fig:twoloop}). In all other
cases we can use a naive anticommuting $\gamma_5$ (NDR scheme), which
avoids spurious finite renormalisations required in the HV
scheme~\cite{Trueman:1995ca}. 

We have performed our calculation in the $\MS$ scheme as described in
Section~\ref{sec:structure}. The renormalisation of masses and
couplings is performed in the usual way. 

\begin{figure}[t]
  \begin{center}
    \includegraphics[width=0.9\textwidth]{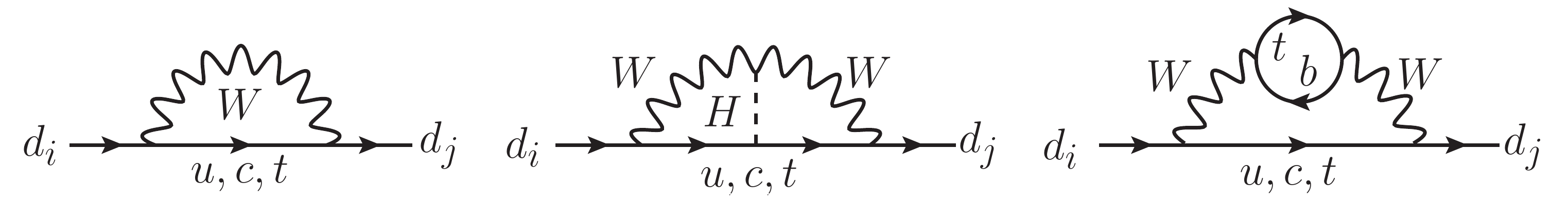}
      \caption{Sample diagrams which imply an
        off-diagonal field renormalisation.
        \label{fig:sdtrans}}
  \end{center}
\end{figure}

In order to ensure the canonical form of the kinetic term for the
down-type quarks, $i \, \bar d_{L,k}\slashed{D}d_{L,j}$, in the
effective theory, we perform a finite off-diagonal field
renormalisation. The exchange of $W$ bosons induces transitions
between quarks of different generations
(cf. Figure~\ref{fig:sdtrans}).
We rediagonalise the kinetic term by including a suitable finite part
in the (matrix-) field renormalisation $Z^{1/2}_{L,ij}$:
\begin{equation}
  d_{L,i}^{\text{bare}}=Z^{1/2}_{L,ij}d_{L,j}, 
  \label{eq:calrenconstants}
\end{equation}
where $i$ denotes the generation of the down-type fermion
($i=1,2,3$). 

The renormalisation leads to a finite result for $X_t^{(EW)}$. As an
additional check we also verified that the full result is analytically
independent of the renormalisation scale $\mu$.


\section{Numerics}\label{sec:numerics}

In this section we present our numerical results and discuss the
theoretical uncertainty of the branching ratios of the rare Kaon
decays. For our numerical analysis we use the central values and
errors of the input parameters given in Table~\ref{tab:num}. As
discussed in detail in Section~\ref{sec:structure}, we use $\alpha$,
$G_F$, and $M_Z$ as the basic input parameters for the electroweak
theory. The mass of the $W$ boson is then not an independent quantity;
we calculate its mass using the approximate formula given in
Reference~\cite{Awramik:2003rn}, which includes the state-of-the-art
higher order corrections.

Converting the on-shell top-quark mass $M_t^{\text{TEV}}$, measured at
Tevatron, to the $\MS$ scheme using three-loop QCD accuracy, we find
$M_t\equiv m_{t}^{\MS, \text{QCD}}(m_t)=163.7\,\text{GeV}$.
For this conversion as well as for the QCD
running of $M_t$ and $\alpha_s$ we use the Mathematica package
RunDec~\cite{Chetyrkin:2000yt}.

\begin{table}[t]
\begin{center}
\begin{tabular}{|c|l|c||c|l|c|}\hline
  \bf{Parameter}               	& \bf{Value}                                    & \bf{Ref.}                     &
  \bf{Parameter}               	& \bf{Value}                                    & \bf{Ref.}                     
  \\\hline
  $\MZ$				& $91.1876(21)$\,GeV				& \cite{Nakamura:2010zzi}	&
  $\alpha_s(\MZ)$		& $0.1184(7)$					& \cite{Nakamura:2010zzi}	
  \\\hline
  $M_H$				& $155(40)$\,GeV				&				&
  $\hat\alpha^{-1}(\MZ)$	& $127.925(16)$					& \cite{Nakamura:2010zzi}
  \\\hline
  $M_t^{\text{TEV}}$	        & $173.3(1.1)$\,GeV				& \cite{:2010yx}		&
  $G_F$				& $1.166\,367(5)\times 10^{-5}\,\text{GeV}^{-2}$& \cite{Nakamura:2010zzi}
  \\\hline
  $m_c(m_c)$			& $1.279(13)$\,GeV				& \cite{Chetyrkin:2009fv}	&
  $\lambda$			& $0.2255(7)$					& \cite{Antonelli:2008jg}
  \\\hline
  $\hat s^2_{\text{ND}}(\MZ)$	& $0.2315(13)$					& \cite{Nakamura:2010zzi}	&
  $\left| V_{cb} \right|$	& $4.06(13) \times 10^{-2}$ 			& \cite{Nakamura:2010zzi}
  \\\hline
  $\kappa_+$			& $0.5173(25)\times 10^{-10}$			& \cite{Mescia:2007kn}		&
  $\bar \rho$			& $0.141^{+0.029}_{-0.017}$			& \cite{Charles:2004jd}
  \\\hline
  $\kappa_L$			& $2.231(13)\times 10^{-10}$			& \cite{Mescia:2007kn}		&
  $\bar \eta$			& $0.343(16)$					& \cite{Charles:2004jd}
  \\\hline
  $\vert\epsilon_K\vert$	& $2.228(11)\times 10^{-3}$			& \cite{Nakamura:2010zzi}	&
  				& 						& 
  \\\hline
\end{tabular}
\caption{Input parameters used in our numerical analysis. \label{tab:num}}
\end{center}
\end{table}

\begin{figure}[h]
\begin{center}
\scalebox{1}{\input{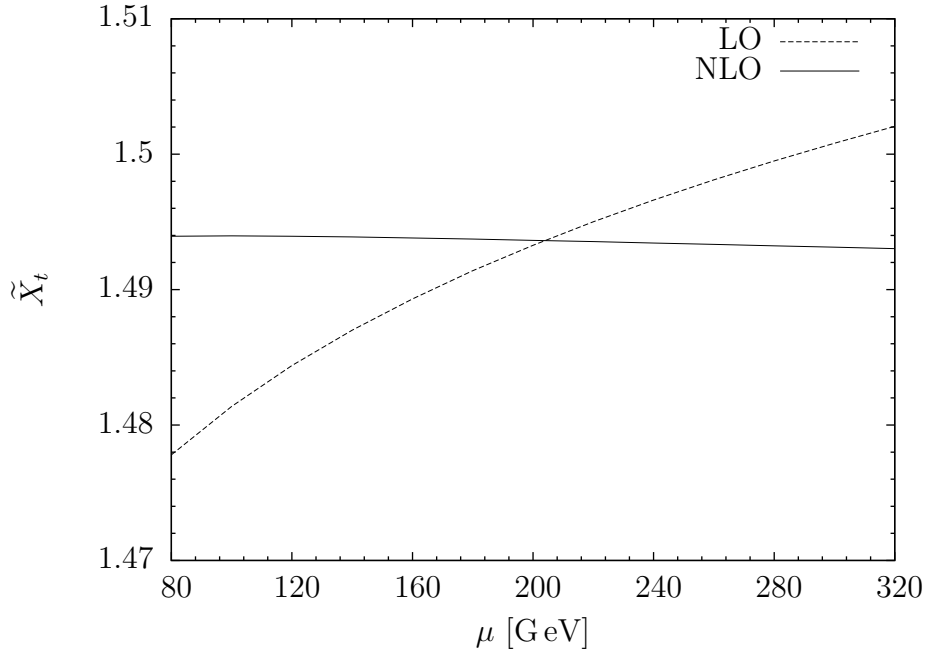}}
\end{center}
\vspace{-5mm}
\caption{$\widetilde X_{t}$ (see text) as a function of $\mu$, for
  $M_H=155\,\text{GeV}$. The LO result is represented by the dashed
  line, the solid line includes the full two-loop electroweak
  corrections, which cancel the $\mu_t$-dependence of the LO result
  almost completely. }
\label{fig:Xtmu}
\end{figure}

The electroweak correction term $X_t^{(EW)}$ cancels the scheme and
scale dependence of the prefactor $\alpha/\sin^2\theta_W$ up to higher
orders in the electroweak interaction. The remaining scheme and scale
dependence will serve as an estimate of the theoretical uncertainty of
our result. To facilitate the discussion, we define the scale and
scheme independent quantity
\begin{equation}
\widetilde X_t =
\frac{\alpha(\mu,M_H)} {\alpha(\mu=M_Z,M_H=155\,\text{GeV})}
\frac{\sin^2\theta_W(\mu=M_Z,M_H=155\,\text{GeV})}
{\sin^2\theta_W(\mu,M_H)} X_{t}(\mu)\, .
\end{equation}
It is formally independent of $\mu$ and coincides with $X_{t}(\mu)$ at
$\mu = M_Z$ and $M_H = 155\,\text{GeV}$. We normalise $\widetilde X_t$
to our central value for the Higgs-boson mass, $M_H=155\,\text{GeV}$; as
we will see below, the dependence on $M_H$ is very weak for
$115\,\text{GeV} < M_H < 200\,\text{GeV}$. The function $\widetilde X_t$
is plotted in Figure~\ref{fig:Xtmu} for $M_H=155\,\text{GeV}$. Here the
dashed line shows the LO result. As is clearly visible, the inclusion
of the two-loop electroweak corrections (solid line) cancels the scale 
dependence of the electroweak input parameters almost completely, up 
to negligible corrections of 0.02\%.

Next we discuss the dependence of our result on the choice of the
renormalisation scheme. The difference between the $\MS$ and on-shell
definition of the parameters $\sin^2\theta_W$ and $m_t^2$, appearing
in the LO effective Hamiltonian, amounts to roughly 4\% and 7\%,
respectively, leading to a large dependence of the branching ratios on
the renormalisation scheme, if the two-loop electroweak corrections
are not included. In turn, we will see that the inclusion of
$X_t^{(EW)}$ cancels this ambiguity almost completely. To get a
quantitative estimate, we evaluate the function $X_t$ numerically in
the three renormalisation schemes described in
Section~\ref{sec:structure}.

\begin{figure}[h]
\begin{center}
\scalebox{1}{\input{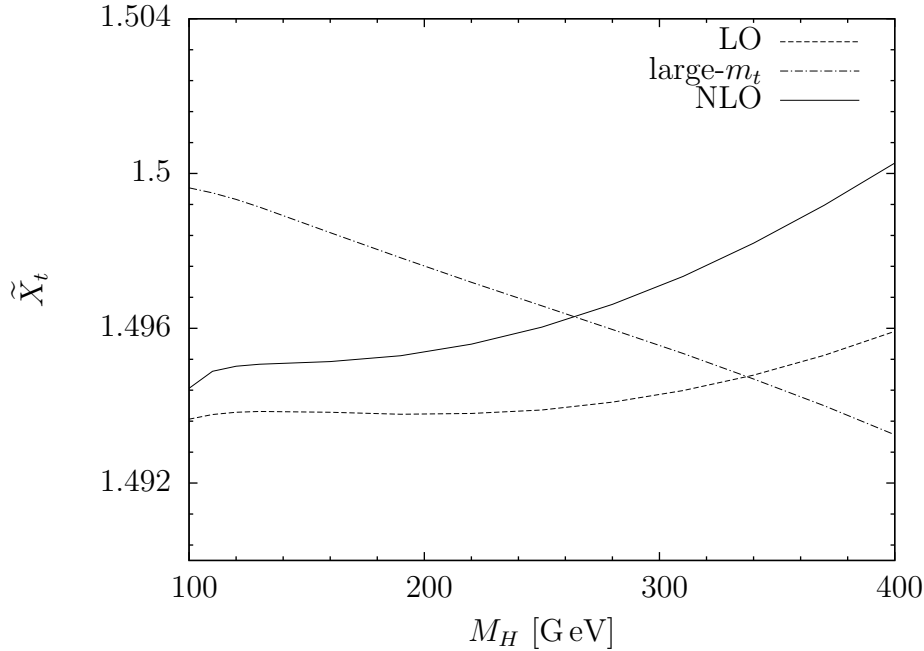}}
\end{center}
\vspace{-5mm}
\caption{$\widetilde X_{t}$ as a function of $M_H$. The LO result is
  represented by the dashed line, the solid line shows the result
  including the full two-loop electroweak corrections. The NLO
  corrections in the limit of large top-quark mass are represented by
  the dashed-dotted line. }
\label{fig:XmhswMS}
\end{figure}

In Figure~\ref{fig:XmhswMS} we show $\widetilde X_t$ in dependence on
the Higgs boson mass $M_H$, where all couplings are defined in the
$\MS$ scheme and all masses in the on-shell scheme. In this scheme the
NLO electroweak corrections are tiny, of the order of one per mil,
even for very large Higgs masses.

\begin{figure}
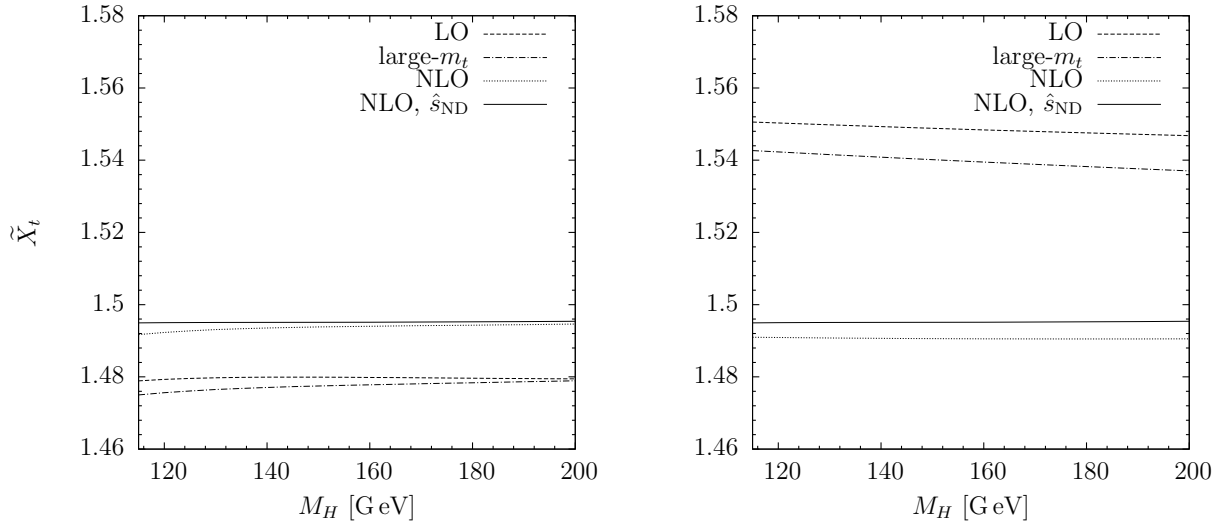

\hspace{-1cm}
\begin{minipage}[t]{0.5\textwidth}
\scalebox{0.8}{\input{X_mht.tex}}
\end{minipage}
\begin{minipage}[t]{0.5\textwidth}
\scalebox{0.8}{\input{X_mh_swOSt.tex}}
\end{minipage}
\caption{$\widetilde X_{t}$ as a function of $M_H$, in two different
  renormalisation schemes. The dashed lines show the LO results, the
  dashed-dotted lines the LO results including the electroweak
  corrections in the large-$m_t$ limit. The full two-loop results are
  represented by the dotted lines. The left panel shows the results
  where all parameters are defined in the $\MS$ scheme. By contrast,
  in the right panel, all parameters apart from $\alpha$ are defined
  in the on-shell scheme. For comparison, we also plot in both panels
  the NLO result, where all masses are defined on-shell and all
  couplings in the $\MS$ scheme. It is represented by the solid
  lines. }
\label{fig:Xtmh}
\end{figure}

In Figure~\ref{fig:Xtmh} we compare the results in two other
schemes. In the left panel we show $\widetilde X_t$, where all
parameters are defined in the $\MS$ scheme. In the right panel, all
parameters are defined in the on-shell scheme, apart from $\alpha$,
which is defined in the $\MS$ scheme. As expected, we observe that for
the on-shell definition of $\sin^2\theta_W$ (right panel) the related
ambiguity is cancelled by a sizeable ($\approx 4\%$) two-loop
correction, whereas for the full $\MS$ definition (left panel) the
electroweak corrections amount to 1\%.

We thus conclude that the on-shell definition of the masses together
with the $\MS$ definition of $\sin^2\theta_W$ is the best choice of
the renormalisation scheme. We can read off the maximal difference of
the three renormalisation schemes from the two NLO curves in
Figure~\ref{fig:Xtmh}, right panel -- it amounts to 0.27\%. For our
numerics below, we will take the average of the two curves and assign
an error of $\pm 0.134\%$ to $X_t$, as resulting from the remaining
uncertainty of the electroweak correction. In total, using the central
values from Table~\ref{tab:num}, we have
\begin{equation} \label{eq:Xtfinal}
X_t = 1.469 \pm 0.017 \pm 0.002 \, ,
\end{equation}
where the first error quantifies the remaining scale uncertainty of
the QCD corrections, and the second error corresponds to the
uncertainty of the electroweak corrections. Here and below, we
determine the QCD error on $X_t$ by varying the scale $\mu_c$ between
80 GeV and 320 GeV. Accordingly, our central value of $X_t$ is the
average of $\max_{\mu}X_t(\mu)$ and $\min_{\mu}X_t(\mu)$, where
$\mu\in [60\,\text{GeV},320\,\text{GeV}]$.

Next, let us comment on the validity of the large-$m_t$ expansion of
the full result, which can be gleaned from Figure~\ref{fig:Xtmh}: It
is now evident that it is always a bad approximation to the full
result, as has actually been expected
before~\cite{Buchalla:1997kz,Gambino:1998rt}.

For convenience we provide an approximate, yet very accurate formula
for the NLO electroweak correction factor $r_X = 1 + X_t^{(EW)}/X_t^{(0)}$:
\begin{equation} \label{eq:ewfit} r_X = 1 - A + B \cdot
  C^{(M_t/165\,\text{GeV})} - D \left( \frac{M_t}{165\,\text{GeV}}
  \right)
\end{equation}
where
\begin{equation}
A = 1.11508\, ,\quad B = 1.12316\, ,\quad C = 1.15338\, ,\quad D = 0.179454\, .
\end{equation}
It approximates the full result within the limits $160\,\text{GeV}
\leq M_t \leq 170\,\text{GeV}$ to an accuracy of better than $\pm
0.05\%$.

Finally, we update the theoretical prediction of the branching ratios,
including the effect of the full two-loop electroweak corrections.
After summation over the three neutrino flavours the resulting
branching ratio for $K^+ \to \pi^+ \nu \bar\nu$ can be written
as\footnote{We have omitted a term which arises from the implicit sum
  over lepton flavours in $P_c$ because it amounts to only 0.2\% of
  the branching ratio. }
\cite{Buchalla:1993wq,Buchalla:1998ba,Isidori:2005xm}
\begin{multline}\label{eq:BR}
  \text{Br} \left(K^+\to\pi^+\nu\bar{\nu}(\gamma)\right) \\ = \kappa_+
  (1+\Delta_{\text{EM}})
  \Bigg[\left(\frac{\text{Im}\lambda_t}{\lambda^5} X_t\right)^2 +
  \left(\frac{\text{Re}\lambda_c}{\lambda} \left(P_c + \delta P_{c,u}
    \right) + \frac{\text{Re}\lambda_t}{\lambda^5} X_t\right)^2
  \Bigg].
\end{multline}
The parameter
\begin{equation}\label{eq:usefulP}
  P_c(X)=\frac{1}{\lambda^4}
  \left(\frac{2}{3}X^e+\frac{1}{3}X^{\tau}\right)
\end{equation}
describes the short-distance contribution of the charm quark, where
$\lambda= \left| V_{us} \right|$, and has been calculated including
electroweak corrections, in Reference~\cite{Brod:2008ss}. The charm
quark contribution of dimension-eight operators at the charm quark
scale $\muc$ \cite{Falk:2000nm} combined with long distance
contributions were calculated in Reference~\cite{Isidori:2005xm} to be
\begin{equation}
  \label{eq:2}
  \delta P_{c,u} = 0.04 \pm 0.02 \, .
\end{equation}

The hadronic matrix element of the low-energy effective Hamiltonian
can be extracted from the well-measured $K_{l3}$ decays, including
isospin breaking and long-distance QED radiative corrections
\cite{Mescia:2007kn,Marciano:1996wy,Bijnens:2007xa}. The long-distance
contributions are contained in the parameters $\kappa_+$, including
NLO and partially NNLO corrections in chiral perturbation
theory. $\Delta_{\text{EM}}$ denotes the long distance QED
corrections~\cite{Mescia:2007kn}.

Including the two-loop electroweak corrections to $X_t$, we find for
the branching ratio of the charged mode
\begin{equation}
\text{Br}(K^+\to\pi^+\nu\bar\nu) = (7.81^{+0.80}_{-0.71}\pm 0.29)
\times 10^{-11}\, , 
\end{equation}
The first error is related to the uncertainties in the input
parameters. The main contributions are ($V_{cb}: 56\%$, $\bar\rho:
21\%$, $m_c: 8\%$, $m_t: 6\%$, $\bar\eta: 4\%$, $\alpha_s: 3\%$,
$\sin^2\theta_W: 1\%$). The second error quantifies the remaining
theoretical uncertainty. In detail, the contributions are ($\delta
P_{c,u}: 46\%$, $X_t(\text{QCD}): 24\%$, $P_c: 20\%$, $\kappa_+:
7\%$, $X_t(\text{EW}): 3\%$), respectively. 

The branching ratio of the $CP$-violating neutral mode involves the
top-quark contribution only and can be written as
\begin{equation}
  \text{Br}\left(K_L\to\pi^0\nu\bar\nu\right)=\kappa_L\left(
    \frac{\text{Im}\lambda_t}{\lambda^5}X_t \right)^2\, .
  \label{eq:brkL}
\end{equation}
Again, the hadronic matrix element can be extracted from the $K_{l3}$
decays and is now parametrised by
$\kappa_L$~\cite{Mescia:2007kn}. There are no more long-distance
contributions, which makes this decay channel exceptionally clean.

Whereas the CP-conserving contribution to the branching ratio is
completely negligible compared to the direct CP-violating contribution
within the Standard Model~\cite{Buchalla:1998ux}, the indirect
CP-violating contribution is of the order of 1\% and should be
included at the current level of accuracy. This can be achieved by
multiplying the branching ratio with the factor~\cite{Buchalla:1996fp}
\begin{equation}
1-\sqrt{2}|\epsilon_K|\frac{1+P_c(X)/(A^2X_t)-\rho}{\eta}\, , 
\end{equation}
where $A=V_{cb}/\lambda^2$, and $\epsilon_K$ describes indirect CP
violation in the neutral Kaon system. Taking this factor into account,
and including again the full two-loop electroweak corrections, we find
\begin{equation}
\text{Br}(K_L\to\pi^0\nu\bar\nu) = (2.43^{+0.40}_{-0.37}\pm 0.06)
\times 10^{-11}\, . 
\end{equation}
The first error is again related to the uncertainties in the input
parameters. Here main contributions are ($V_{cb}: 54\%$, $\bar\eta:
39\%$, $m_t: 6\%$). The contributions to the second, theoretical
uncertainty are ($X_t(\text{QCD}): 73\%$, $\kappa_L: 18\%$,
$X_t(\text{EW}): 8\%$, $\delta P_{c,u}: 1\%$), respectively. All
errors have been added in quadrature.


\section{Conclusions}\label{sec:conclusions}

In this paper, we have calculated the complete two-loop electroweak
matching corrections to $X_t$, the top-quark contribution to the rare
decays $K_L\to\pi^0\nu\bar\nu$, $K^+\to\pi^+\nu\bar\nu$, and $B\to
X_{d,s}\nu\bar\nu$. This is in particular important for rare kaon
decays: future proposals aim at an experimental accuracy of 3\% for
the branching ratios, while the leading order electroweak scheme
ambiguity is of similar size.  Our calculation reduces the scheme
ambiguity in $X_t$ from $\pm 2 \%$ to $\pm 0.134 \%$. The resulting
theory uncertainty in the branching ratios is rendered from dominant
to negligible.

The absolute corrections are small in a renormalisation scheme where
on-shell masses and $\overline{\textrm{MS}}$ coupling constants are
used for the electroweak sector. In addition, we analyse the
convergence in the $\overline{\textrm{MS}}$ scheme and the on-shell
scheme to estimate the remaining perturbative uncertainty.

Our analytic results are summarised by an approximate, but very
accurate formula. We also give the leading term in a small $\sin
\theta_W$ expansion. The full expression can be obtained upon request
from the authors.

\phantomsection
\addcontentsline{toc}{section}{Acknowledgements}
\section*{Acknowledgements}

We would like to thank Gerhard Buchalla, Andrzej Buras, and Paolo
Gambino for useful discussions and comments. 



\appendix

\section{\texorpdfstring{$\boldsymbol{\sin\theta}_{\text{W}}$ Expansion}
			{sin(theta\_W) Expansion}}\label{app:swexpansion}

\begin{figure}[h]
\begin{center}
\scalebox{1}{\input{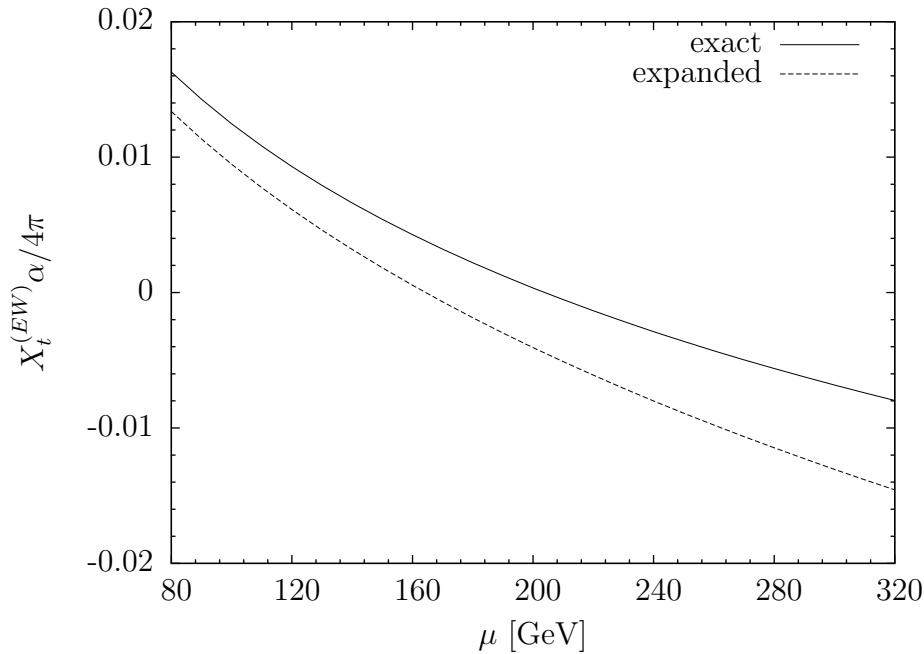}}
\end{center}
\vspace{-5mm}
\caption{The $\sin\theta_W$ expansion of $X_{t}^{(EW)} \alpha / 4\pi$
  in the $\MS$ scheme as a function of the renormalisation scale
  $\mu$. The solid line shows the full result, while the dashed line
  corresponds to the leading term of the expansion. }
\label{fig:swexp}
\end{figure}

The explicit expression of the full two-loop electroweak correction
$X_{t}^{(EW)}$ is too long to be given explicitly here. The result
significantly simplifies if we expand in the small
parameter $\sin \theta_W$ - see Figure~\ref{fig:swexp} for the validity of the
expansion.

In the $\overline{\textrm{MS}}$ scheme, after
normalising the effective Hamiltonian to $G_F$, we find
\begin{align}
  \begin{split}
    X^{(EW)}(\xt,a,\hat s_\text{ND},\mu) &= \frac{1}{128\hat s_\text{ND}^2}
    \left( \sum_{i=1}^{17} c_i A_i + \mathcal{O}\left( \hat s_\text{ND} \right) \right)\, ,
  \end{split}
  \label{eq:xtopswexpansion}
\end{align}
where $a=\big(M_H/m_t^{\MS}\big)^2$,
\begin{flalign*}
c_1\:\,	& = \frac{1}{3 a (\xt-1)^2 \xt} 									\, ,&	
c_2\:\,	& = \frac{1}{(\xt-1)^3 (a \xt-1)} 			\varphi_1(\tfrac{1}{4})				\, ,\\
c_3\:\,	& = \frac{1}{2 (\xt-1)^3 (a \xt-1)} 			\varphi_1(\tfrac{a}{4})				\, ,&
c_4\:\,	& = \frac{1}{2 (\xt-1)^3 (a \xt-1)} 			\varphi_1(\tfrac{1}{4\xt})			\, ,\\
c_5\:\,	& = \frac{1}{2 (\xt-1)^3 (a \xt-1)}			\varphi_1(\tfrac{\xt}{4}) 			\, ,&
c_6\:\,	& = \frac{1}{(\xt-1)^3 (a \xt-1)} 			\varphi_1(\tfrac{a\xt}{4})			\, ,\\
c_7\:\,	& = \frac{1}{2 a^2 \xt^2 (\xt-1)^3 (a \xt-1)} 		\varphi_2(\tfrac{1}{a\xt},\tfrac{1}{a})		\, ,&
c_8\:\,	& = \frac{1}{a \xt-1} 					\log^2(a)					\, ,\\
c_9\:\,	& = \frac{1}{3 (\xt-1)^3 (a \xt-1)} 			\log(\xt)					\, ,&
c_{10}	& = \frac{1}{2 a (\xt-1)^4 \xt (a \xt-1)} 		\log^2(\xt)					\, ,\\
c_{11} 	& = \frac{1}{(\xt-1)^2}					\log(\tfrac{\mu^2}{M_{\text{W}}^2})		\, ,& 
c_{12}	& = \frac{1}{(\xt-1)^3}					\log(\xt)\log(\tfrac{\mu^2}{M_{\text{W}}^2})	\, ,\\
c_{13} 	& = \frac{1}{(\xt-1)^2 (a \xt-1)}			\log(a)		      				\, ,&
c_{14}	& = \frac{1}{2 a (\xt-1)^3 \xt (a \xt-1)}		\log(\xt)\log(a)      				\, ,\\
c_{15}	& = \frac{1}{(\xt-1)^2} 				\text{Li}_2( 1 - a)   				\, ,&
c_{16}	& = \frac{1}{a \xt} 					\text{Li}_2( 1 - \xt) 				\, ,\\	
c_{17}	& = \frac{1}{a (\xt-1)^2 \xt} 				\text{Li}_2( 1 - a\xt)
\end{flalign*}
and
\begin{align*}
A_1 = &		+(16 - 48 a) \pi^2
		+(288 a - (32 - 88 a) \pi^2) \xt
		+(2003 a +4 (4 - 6 a - a^2) \pi^2) \xt^2				\\&
		+(9 a (93 + 28 a) - 4 a (3 - 2 a + 8 a^2) \pi^2) \xt^3			\\&
		+(3 a (172 - 49 a - 32 a^2) + 4 a (20 - a + 16 a^2) \pi^2) \xt^4	\\&
		- (3 a (168 + 11 a - 24 a^2) + 4 a (45 + 8 a^2) \pi^2) \xt^5		\\&
		+96 a \pi^2 \xt^6\, ,
\displaybreak[0]\\[1em]
A_2 = &		-768 \xt
		-(525 - 867 a) \xt^2
		+(303 + 318 a) \xt^3
		-195 a \xt^4\, ,
\displaybreak[0]\\[1em]
A_3 = &	 	-8 (95 - 67 a + 11 a^2) \xt^2
		+2 (662 - 78 a - 177 a^2 + 40 a^3) \xt^3		\\&
		-(608 + 476 a - 595 a^2 + 114 a^3) \xt^4
		+(44 + 188 a - 321 a^2 + 103 a^3 - 8 a^4) \xt^5		\\&
		-a (28 - 72 a + 33 a^2 - 4 a^3) \xt^6\, ,
\displaybreak[0]\\[1em]
A_4 = &		+48
		-10 (57 + 4 a) \xt
		+51 (29 + 10 a) \xt^2
		-(841 + 1265 a) \xt^3
		+(308 + 347 a) \xt^4	\\&
		-(28 - 40 a) \xt^5
		+12 a \xt^6\, ,
\displaybreak[0]\\[1em]
A_5 = &		+768
		+(816 - 768 a) \xt
		+(1240 - 1232 a) \xt^2
		-4 (415 + 2 a) \xt^3
		+(311 + 722 a) \xt^4	\\&
		+(145 - 267 a) \xt^5
		-(36 + 51 a) \xt^6
		+ 20 a \xt^7\, ,
\displaybreak[0]\\[1em]
A_6 = &		+328 \xt
		-(536 + 900 a) \xt^2
		+(208 + 1584 a + 670 a^2) \xt^3
		-a (668 + 1161 a + 225 a^2) \xt^4	\\&
		+a^2 (479 + 362 a + 28 a^2) \xt^5
		-a^3 (143 + 42 a) \xt^6
		+16 a^4 \xt^7\, ,
\displaybreak[0]\\[1em]
A_7 = &		+32
		-4 (44 - 9 a) \xt
		+(384 - 322 a - 400 a^2) \xt^2
		-(400 - 869 a - 1126 a^2 - 696 a^3) \xt^3				\\&
		+ 2 (80 - 488 a - 517 a^2 - 631 a^3 - 264 a^4) \xt^4			\\&
		+(48 + 394 a + 269 a^2 + 190 a^3 + 882 a^4 + 196 a^5) \xt^5		\\&
		-(64 - 58 a - 89 a^2 - 95 a^3 + 34 a^4 + 296 a^5 + 32 a^6) \xt^6	\\&
		+(16 - 59 a - 79 a^2 + 256 a^3 - 239 a^4 + 57 a^5 + 48 a^6) \xt^7	
		+(1 - a)^3 a^2 (29 + 16 a) \xt^8\, ,
\displaybreak[0]\\[1em]
A_8 = &	 	+28 a^2 \xt^2 
		-32 a^3 \xt^3\, ,
\displaybreak[0]\\[1em]
A_9 = &	 	-288
		+36 (1 + 8 a) \xt
		+6 (647 + 87 a) \xt^2
		+5 (55 - 927 a - 132 a^2) \xt^3			\\&
		-(1233 + 98 a - 879 a^2 - 192 a^3) \xt^4
		+(360 + 1371 a - 315 a^2 - 264 a^3) \xt^5	\\&
		-24 a (17 - 4 a^2) \xt^6\, ,
\displaybreak[0]\\[1em]
A_{10}= &	+32 
		+ 4 (-44 + 29 a) \xt 
		- 12 (-32 + 77 a + 31 a^2) \xt^2				\\& 
		+ 2 (-200 + 837 a + 767 a^2 + 182 a^3) \xt^3 
		- 2 (-80 + 625 a + 905 a^2 + 520 a^3 + 82 a^4) \xt^4		\\& 
		+ (48 + 1079 a + 590 a^2 + 1002 a^3 + 462 a^4 + 32 a^5) \xt^5	\\& 
		+ (-64 - 1160 a - 501 a^2 - 364 a^3 - 486 a^4 - 72 a^5) \xt^6	\\& 
		+ (16 + 729 a + 1038 a^2 + 38 a^3 + 238 a^4 + 52 a^5) \xt^7	\\& 
		- a (192 + 743 a + 50 a^3 + 12 a^4) \xt^8 + 192 a^2 \xt^9\, ,
\displaybreak[0]\\[1em]
A_{11} = &	 +16 \xt
		 +324 \xt^2
		 -36 \xt^4\, ,
\displaybreak[0]\\[1em]
A_{12}= &	 +216 \xt
		 -672 \xt^2
		 +152 \xt^3\, ,
\displaybreak[0]\\[1em]
A_{13} = &	 -16 \xt
		 +(16 - 42 a) \xt^2
		 +(16 + 21 a + 60 a^2) \xt^3		\\&
		 -(16 - 21 a + 45 a^2 + 32 a^3) \xt^4
		 -a^2 (7 - 24 a) \xt^5\, ,
\displaybreak[0]\\[1em]
A_{14}= & 	-32 
		+ (144 - 68 a) \xt 
		+ (-240 + 334 a + 332 a^2) \xt^2 
		+ (160 - 551 a - 660 a^2 - 364 a^3) \xt^3			\\& 
		+ a (329 + 451 a + 650 a^2 + 164 a^3) \xt^4 
		+ (-48 - a - 59 a^2 - 523 a^3 - 316 a^4 - 32 a^5) \xt^5		\\& 
		+ (16 - 43 a - 93 a^2 + 255 a^3 + 287 a^4 + 32 a^5) \xt^6 
		- a^2 (-29 + 42 a + 103 a^2 + 8 a^3) \xt^7\, ,
\displaybreak[0]\\[1em]
A_{15}= &	-144 (1 - a)^2 \xt^2
		+144 (1 - a)^2 \xt^3
		-36 (1 - a)^2 \xt^4\, ,
\displaybreak[0]\\[1em]
A_{16}= &	-32 + 96 a
		+(48 - 32 a) \xt
		-176 a \xt^2
		-(16 - 74 a) \xt^3
		+212 a \xt^4\, ,
\displaybreak[0]\\[1em]
A_{17}= &	-32
		+(64 - 100 a) \xt
	-8 (4 - 34 a - 29 a^2) \xt^2
		-4 a (34 + 170 a + 33 a^2) \xt^3	\\&
		+8 a^2 (47 + 51 a + 4 a^2) \xt^4
		-16 a^3 (15 + 4 a) \xt^5
		+32 a^4 \xt^6 \, . 
\end{align*}
Here we use
\begin{equation*}
  \text{Li}_2(\xi)=-\int_0^1\!\frac{\log(1-\xi t)}{t}dt \, , 
\end{equation*}
and the two-loop functions $\varphi_1$ and $\varphi_2$ are given
by~\cite{Davydychev:1992mt}
\begin{equation}
\begin{split}
\varphi_1(z) = 
\begin{cases}
  4\sqrt{\frac{z}{1-z}}\text{Cl}_2(2\arcsin(\sqrt{z})) \, , \quad
  0\leq z< 1\, ,
  \\[2mm]
  \frac{1}{\lambda_z}\left(2\ln^2\frac{1-\lambda_z}{2} -
    4\text{Li}_2\frac{1-\lambda_z}{2} - \ln^2(4z) +
    \frac{1}{3}\pi^2\right)\, , \quad z>1\, ,
\end{cases}
\end{split}
\end{equation}
and 
\begin{equation}
\begin{split}
\varphi_2(x,y) = 
\begin{cases}
  \frac{1}{\lambda} \big\{\frac{\pi^2}{3} + 2\ln\left(
    \frac{1}{2}(1+x-y-\lambda) \right)
  \ln\left( \frac{1}{2}(1-x+y-\lambda) \right) - \ln x \ln y \\[1mm]
  -2\text{Li}_2\left( \frac{1}{2}(1+x-y-\lambda) \right) -2\text{Li}_2
  \left( \frac{1}{2}(1-x+y-\lambda) \right)\big\} \, , \lambda^2\geq 0\, ,
  \sqrt{x}+\sqrt{y} \leq 1\, ,
  \\[3mm]
  \frac{2}{\sqrt{-\lambda^2}} \bigg\{ \text{Cl}_2 \left( 2 \arccos
    \left( \frac{-1+x+y}{2\sqrt{xy}} \right)\right) + \text{Cl}_2
  \left( 2 \arccos \left( \frac{1+x-y}{2\sqrt{x}} \right)\right) \\ +
  \text{Cl}_2 \left( 2 \arccos \left( \frac{1-x+y}{2\sqrt{y}}
    \right)\right)\bigg\}\, , \quad \lambda^2\leq 0\, ,
  \sqrt{x}+\sqrt{y} \geq 1\, .
\end{cases}
\end{split}
\end{equation}
Here $\lambda_z=\sqrt{1-1/z}$ and $\lambda = \sqrt{(1-x-y)^2 -
  4xy}$. The Clausen function is defined by $\text{Cl}_2(z) =
-\int_0^\theta d\theta \ln|2\sin(\theta/2)|$.
 
\boldmath
\section{The large-$m_t$ Expansion}
\unboldmath

The two-loop electroweak corrections to the $bbZ$ vertex, denoted by
$\tau_b^{(2)}$, have been calculated in the limit of a large top-quark
mass by Barbieri et al. in~\cite{Barbieri:1992dq,Barbieri:1992nz} and
were confirmed by Fleischer, Tarasov, and
Jegerlehner~\cite{Fleischer:1994cb}, who found a particularly simple
analytic form of the results. Buchalla and Buras have extracted from
this result the corrections to the $sdZ$ vertex, which they used for
their analysis of the $K\to\pi\nu\bar\nu$ decays
in~\cite{Buchalla:1997kz}. We will now take the limit $m_t \to \infty$
in our complete result for the $sd\nu\nu$ transition and compare it
with the result in~\cite{Fleischer:1994cb}.

Several important points should be mentioned here: As observed
in~\cite{Buchalla:1997kz}, only $Z$ penguin diagrams contribute to the
$sd\nu\nu$ transition in the large-$m_t$ limit. The results
in~\cite{Fleischer:1994cb} have been obtained in the so-called
``gaugeless limit'', where in particular the $W$ boson field does not
appear. Accordingly, the parameter corresponding to our $x_t$ is
defined in~\cite{Fleischer:1994cb} by $x_t \equiv \sqrt{2}G_\mu
m_t^2/(16\pi^2)$ and will be denoted by $\tilde x_t$ in our paper. As
a consequence, the result for $\tau_b^{(2)}$ is normalised to $G_F^2$
in~\cite{Fleischer:1994cb}.  On the other hand, we performed a full
Standard Model calculation and afterwards took the limit $m_t \to
\infty$.

Thus we now take the large-$m_t$ expansion of our result, factor out
$G_F^2$, and perform a finite renormalisation of the top-quark mass in
our LO result, by replacing $m_t^{\MS} = M_t + \delta M_t$, in order
to transform into the on-shell scheme. Here $\delta M_t$ is given in
the large-$m_t$ limit by
\begin{equation}
  \frac{\delta M_t}{M_t} = \frac{e^2}{16\pi^2 s_W^2} x_t \left( \frac{3}{a} + 1 -
    \frac{1}{2}a - \frac{1}{16} (4 a^{1/2} - a^{3/2}) g(a) +
    \frac{1}{16} a^2 \log a\right) \, ,
\end{equation}
and
\begin{equation}
g(a) = 2\sqrt{a-4} \left[ \text{arctanh} \left(
    \frac{2-a}{\sqrt{(a-4)a}} \right) + \text{arctanh} \left(
    \sqrt{\frac{a}{a-4}} \right) \right] \, .
\end{equation}
In this way we reproduce the result
in~\cite{Fleischer:1994cb}:
\begin{equation}
\begin{split}
  \tau_b^{(2),\text{on-shell}} & = 9 -\frac{13}{4}a -2 a^2 -
  \left(\frac{1}{24} + \frac{7}{12} a^2 -\frac{1}{2} a^3 \right) \pi^2
  - \left( \frac{19}{4} a + \frac{3}{2} a^2\right) \ln a
  \\
 & - \left( \frac{7}{4} a^2 - \frac{3}{2} a^3\right) \ln^2
 a-\left(\frac{7}{4} -\frac{15}{2}a + \frac{39}{4}a^2 -4a^3\right)
  \text{Li}_2(1-a) -\left(2-\frac{a}{2}\right)\sqrt{a}g(a) \\
 & -\frac{1}{2}\left( 7 - 18a + \frac{33}{4}a^2 - a^3\right)
 \varphi_1\left( \frac{a}{4}\right)\, .
\end{split}
\end{equation}
It corresponds to the effective Hamiltonian in the limit of large
top-quark mass
\begin{equation}
\label{eq:Hefflt}
\mathcal{H}_{\text{eff}} = \frac{4G_F}{\sqrt{2}}
\frac{\alpha}{2\pi\sin^2\theta_{\text{W}}} \lambda_t \left(
  \frac{\xt}{8} + \frac{\alpha}{4\pi}
  \frac{x_t^2}{32\sin^2\theta_{\text{W}}} 
  (3+\tau_b^{(2)}) \right) Q_\nu \, .
\end{equation}
Our result in the $\MS$ scheme is given by
\begin{equation}
\begin{split}
   \tau_b^{(2),\MS}&= -2 -\frac{11}{4}a -2 a^2 -
  \left(\frac{1}{24} + \frac{7}{12} a^2 -\frac{a^3}{2} \right) \pi^2
  \\
  & - \left( \frac{7}{4} a + 2 a^2\right) \ln a- \left( \frac{7}{4}
    a^2
    - \frac{3}{2} a^3\right) \ln^2 a \\
  &-\left(\frac{7}{4} -\frac{15}{2}a + \frac{39}{4}a^2 -4a^3\right)
  \text{Li}_2(1-a) -\frac{1}{2}\left( 7 - 18a + \frac{33}{4}a^2 - a^3\right)
 \varphi_1\left( \frac{a}{4}\right)
\end{split}
\end{equation}
for $\mu_t = M_t$. It is normalised to $G_F$ and thus independent of
the tadpole contribution.


\phantomsection
\addcontentsline{toc}{section}{References}
\bibliography{kpinunu_Xt}

\begin{thebibliography}{10}
\newcommand{\enquote}[1]{``#1''}
\providecommand{\url}[1]{\texttt{#1}}
\providecommand{\urlprefix}{URL }
\expandafter\ifx\csname urlstyle\endcsname\relax
  \providecommand{\doi}[1]{doi:\discretionary{}{}{}#1}\else
  \providecommand{\doi}{doi:\discretionary{}{}{}\begingroup
  \urlstyle{rm}\Url}\fi
\providecommand{\eprint}[2][]{\url{#2}}

\bibitem{Buchalla:1993wq}
G.~Buchalla and A.~J. Buras.
\newblock \enquote{{The rare decays $K^+ \to \pi^+ \nu \bar \nu$ and $K_L \to
  \mu^+ \mu^-$ beyond leading logarithms}}.
\newblock Nucl. Phys. \textbf{B412}:106, 1994.
\newblock \doi{10.1016/0550-3213(94)90496-0}.
\newblock \eprint{hep-ph/9308272}.

\bibitem{Buchalla:1998ba}
G.~Buchalla and A.~J. Buras.
\newblock \enquote{{The rare decays $K \to \pi \nu \bar \nu$, $B \to X \nu
  \bar\nu$ and $B \to l^+ l^-$: An update}}.
\newblock Nucl. Phys. \textbf{B548}:309, 1999.
\newblock \doi{10.1016/S0550-3213(99)00149-2}.
\newblock \eprint{hep-ph/9901288}.

\bibitem{Gorbahn:2004my}
M.~Gorbahn and U.~Haisch.
\newblock \enquote{{Effective Hamiltonian for non-leptonic $|\Delta F| = 1$
  decays at NNLO in QCD}}.
\newblock Nucl. Phys. \textbf{B713}:291, 2005.
\newblock \doi{10.1016/j.nuclphysb.2005.01.047}.
\newblock \eprint{hep-ph/0411071}.

\bibitem{Buras:2005gr}
A.~J. Buras, M.~Gorbahn, U.~Haisch and U.~Nierste.
\newblock \enquote{{The rare decay $K^+ \to \pi^+ \nu \bar\nu$ at the
  next-to-next- to-leading order in QCD}}.
\newblock Phys. Rev. Lett. \textbf{95}:261805, 2005.
\newblock \doi{10.1103/PhysRevLett.95.261805}.
\newblock \eprint{hep-ph/0508165}.

\bibitem{Buras:2006gb}
A.~J. Buras, M.~Gorbahn, U.~Haisch and U.~Nierste.
\newblock \enquote{{Charm quark contribution to $K^+ \to \pi^+ \nu \bar\nu$ at
  next-to-next-to-leading order}}.
\newblock JHEP \textbf{11}:002, 2006.
\newblock \eprint{hep-ph/0603079}.

\bibitem{Brod:2008ss}
J.~Brod and M.~Gorbahn.
\newblock \enquote{{Electroweak Corrections to the Charm Quark Contribution to
  $K^+ \to \pi^+ \nu \bar\nu$}}.
\newblock Phys. Rev. \textbf{D78}:034006, 2008.
\newblock \doi{10.1103/PhysRevD.78.034006}.
\newblock \eprint{0805.4119}.

\bibitem{Buchalla:1992zm}
G.~Buchalla and A.~J. Buras.
\newblock \enquote{{QCD corrections to the $\bar s\, d\, Z$ vertex for
  arbitrary top quark mass}}.
\newblock Nucl. Phys. \textbf{B398}:285, 1993.
\newblock \doi{10.1016/0550-3213(93)90110-B}.

\bibitem{Misiak:1999yg}
M.~Misiak and J.~Urban.
\newblock \enquote{{QCD corrections to FCNC decays mediated by Z-penguins and
  W-boxes}}.
\newblock Phys. Lett. \textbf{B451}:161, 1999.
\newblock \doi{10.1016/S0370-2693(99)00150-1}.
\newblock \eprint{hep-ph/9901278}.

\bibitem{Buchalla:1997kz}
G.~Buchalla and A.~J. Buras.
\newblock \enquote{{Two-loop large-$m_t$ electroweak corrections to $K
  \rightarrow\pi\nu\bar \nu$ for arbitrary Higgs boson mass}}.
\newblock Phys. Rev. \textbf{D57}:216, 1998.
\newblock \doi{10.1103/PhysRevD.57.216}.
\newblock \eprint{hep-ph/9707243}.

\bibitem{Inami:1980fz}
T.~Inami and C.~S. Lim.
\newblock \enquote{{Effects of Superheavy Quarks and Leptons in Low-Energy Weak
  Processes $K_L \rightarrow \mu \bar\mu$, $K^+\rightarrow \pi^+\nu\bar\nu$ and
  $K^0 \leftrightarrow\overline{K^0}$}}.
\newblock Prog. Theor. Phys. \textbf{65}:297, 1981 [Erratum-ibid. \textbf{65}:1772, 1981].
\newblock \doi{10.1143/PTP.65.297}.

\bibitem{Marciano:1988vm}
W.~J. Marciano and A.~Sirlin.
\newblock \enquote{{Electroweak Radiative Corrections to $\tau$ Decay}}.
\newblock Phys. Rev. Lett. \textbf{61}:1815, 1988.
\newblock \doi{10.1103/PhysRevLett.61.1815}.

\bibitem{vanRitbergen:1999fi}
T.~van Ritbergen and R.~G. Stuart.
\newblock \enquote{{On the precise determination of the Fermi coupling constant
  from the muon lifetime}}.
\newblock Nucl. Phys. \textbf{B564}:343, 2000.
\newblock \doi{10.1016/S0550-3213(99)00572-6}.
\newblock \eprint{hep-ph/9904240}.

\bibitem{Jegerlehner:2001fb}
F.~Jegerlehner, M.~Y. Kalmykov and O.~Veretin.
\newblock \enquote{{$\overline{\text{MS}}$ vs. pole masses of gauge bosons:
  Electroweak bosonic two-loop corrections}}.
\newblock Nucl. Phys. \textbf{B641}:285, 2002.
\newblock \doi{10.1016/S0550-3213(02)00613-2}.
\newblock \eprint{hep-ph/0105304}.

\bibitem{Jegerlehner:2002em}
F.~Jegerlehner, M.~Y. Kalmykov and O.~Veretin.
\newblock \enquote{{$\overline{\text{MS}}$ vs pole masses of gauge bosons. II:
  Two-loop electroweak fermion corrections}}.
\newblock Nucl. Phys. \textbf{B658}:49, 2003.
\newblock \doi{10.1016/S0550-3213(03)00177-9}.
\newblock \eprint{hep-ph/0212319}.

\bibitem{Degrassi:1996ps}
G.~Degrassi, P.~Gambino and A.~Sirlin.
\newblock \enquote{{Precise calculation of $M_W$,
  $\sin^2\theta^{\overline{\text{MS}}}$, and $\sin^2\theta_{\text{eff}}$}}.
\newblock Phys. Lett. \textbf{B394}:188, 1997.
\newblock \doi{10.1016/S0370-2693(96)01677-2}.
\newblock \eprint{hep-ph/9611363}.

\bibitem{Bobeth:1999mk}
C.~Bobeth, M.~Misiak and J.~Urban.
\newblock \enquote{{Photonic penguins at two loops and $m_t$-dependence of
  $\text{BR}(B\rightarrow X_s l^+ l^-)$}}.
\newblock Nucl. Phys. \textbf{B574}:291, 2000.
\newblock \doi{10.1016/S0550-3213(00)00007-9}.
\newblock \eprint{hep-ph/9910220}.

\bibitem{Davydychev:1992mt}
A.~I. Davydychev and J.~B. Tausk.
\newblock \enquote{{Two loop selfenergy diagrams with different masses and the
  momentum expansion}}.
\newblock Nucl. Phys. \textbf{B397}:123, 1993.
\newblock \doi{10.1016/0550-3213(93)90338-P}.

\bibitem{Hahn:2000kx}
T.~Hahn.
\newblock \enquote{{Generating Feynman diagrams and amplitudes with FeynArts
  3}}.
\newblock Comput. Phys. Commun. \textbf{140}:418, 2001.
\newblock \doi{10.1016/S0010-4655(01)00290-9}.
\newblock \eprint{hep-ph/0012260}.

\bibitem{Vermaseren:2000nd}
J.~A.~M. Vermaseren.
\newblock \enquote{{New features of FORM}} 2000.
\newblock \eprint{math-ph/0010025}.

\bibitem{Trueman:1995ca}
T.~L. Trueman.
\newblock \enquote{{Spurious anomalies in dimensional renormalization}}.
\newblock Z. Phys. \textbf{C69}:525, 1996.
\newblock \doi{10.1007/s002880050057}.
\newblock \eprint{hep-ph/9504315}.

\bibitem{Awramik:2003rn}
M.~Awramik, M.~Czakon, A.~Freitas and G.~Weiglein.
\newblock \enquote{{Precise prediction for the W-boson mass in the standard
  model}}.
\newblock Phys. Rev. \textbf{D69}:053006, 2004.
\newblock \doi{10.1103/PhysRevD.69.053006}.
\newblock \eprint{hep-ph/0311148}.

\bibitem{Chetyrkin:2000yt}
K.~G. Chetyrkin, J.~H. Kuhn and M.~Steinhauser.
\newblock \enquote{{RunDec: A Mathematica package for running and decoupling of
  the strong coupling and quark masses}}.
\newblock Comput. Phys. Commun. \textbf{133}:43, 2000.
\newblock \doi{10.1016/S0010-4655(00)00155-7}.
\newblock \eprint{hep-ph/0004189}.

\bibitem{Nakamura:2010zzi}
K.~Nakamura.
\newblock \enquote{{Review of particle physics}}.
\newblock J. Phys. \textbf{G37}:075021, 2010.
\newblock \doi{10.1088/0954-3899/37/7A/075021}.

\bibitem{:2010yx}
{Tevatron Electroweak Working Group}.
\newblock \enquote{{Combination of CDF and D0 Results on the Mass of the Top
  Quark}} 2010.
\newblock \eprint{1007.3178}.

\bibitem{Chetyrkin:2009fv}
K.~G. Chetyrkin \emph{et~al.}
\newblock \enquote{{Charm and Bottom Quark Masses: an Update}}.
\newblock Phys. Rev. \textbf{D80}:074010, 2009.
\newblock \doi{10.1103/PhysRevD.80.074010}.
\newblock \eprint{0907.2110}.

\bibitem{Antonelli:2008jg}
M.~Antonelli \emph{et~al.}
\newblock \enquote{{Precision tests of the Standard Model with leptonic and
  semileptonic kaon decays}} 2008.
\newblock \eprint{0801.1817}.

\bibitem{Mescia:2007kn}
F.~Mescia and C.~Smith.
\newblock \enquote{{Improved estimates of rare K decay matrix-elements from
  $K_{\ell 3}$ decays}}.
\newblock Phys. Rev. \textbf{D76}:034017, 2007.
\newblock \doi{10.1103/PhysRevD.76.034017}.
\newblock \eprint{0705.2025}.

\bibitem{Charles:2004jd}
J.~Charles \emph{et~al.}
\newblock \enquote{{CP violation and the CKM matrix: Assessing the impact of
  the asymmetric $B$ factories}}.
\newblock Eur. Phys. J. \textbf{C41}:1, 2005.
\newblock \doi{10.1140/epjc/s2005-02169-1}.
\newblock \eprint{hep-ph/0406184}.

\bibitem{Gambino:1998rt}
P.~Gambino, A.~Kwiatkowski and N.~Pott.
\newblock \enquote{{Electroweak effects in the $B^0-\overline{B^0}$ mixing}}.
\newblock Nucl. Phys. \textbf{B544}:532, 1999.
\newblock \doi{10.1016/S0550-3213(98)00860-8}.
\newblock \eprint{hep-ph/9810400}.

\bibitem{Isidori:2005xm}
G.~Isidori, F.~Mescia and C.~Smith.
\newblock \enquote{{Light-quark loops in $K \to \pi \nu \bar\nu$}}.
\newblock Nucl. Phys. \textbf{B718}:319, 2005.
\newblock \doi{10.1016/j.nuclphysb.2005.04.008}.
\newblock \eprint{hep-ph/0503107}.

\bibitem{Falk:2000nm}
A.~F. Falk, A.~Lewandowski and A.~A. Petrov.
\newblock \enquote{{Effects from the charm scale in $K^+\rightarrow
  \pi^+\nu\bar\nu$}}.
\newblock Phys. Lett. \textbf{B505}:107, 2001.
\newblock \doi{10.1016/S0370-2693(01)00343-4}.
\newblock \eprint{hep-ph/0012099}.

\bibitem{Marciano:1996wy}
W.~J. Marciano and Z.~Parsa.
\newblock \enquote{{Rare kaon decays with 'missing energy'}}.
\newblock Phys. Rev. \textbf{D53}:1, 1996.
\newblock \doi{10.1103/PhysRevD.53.R1}.

\bibitem{Bijnens:2007xa}
J.~Bijnens and K.~Ghorbani.
\newblock \enquote{{Isospin breaking in $K\pi$ vector form-factors for the weak
  and rare decays $K_{\ell3}$, $K\to\pi\nu\bar\nu$ and $K\to\pi\ell^+\ell^-$}}
  2007.
\newblock \eprint{0711.0148}.

\bibitem{Buchalla:1998ux}
G.~Buchalla and G.~Isidori.
\newblock \enquote{{The CP conserving contribution to
  $K_L\rightarrow\pi^0\nu\bar\nu$ in the standard model}}.
\newblock Phys. Lett. \textbf{B440}:170, 1998.
\newblock \doi{10.1016/S0370-2693(98)01088-0}.
\newblock \eprint{hep-ph/9806501}.

\bibitem{Buchalla:1996fp}
G.~Buchalla and A.~J. Buras.
\newblock \enquote{{$K \to \pi \nu \bar \nu$ and high precision determinations
  of the CKM matrix}}.
\newblock Phys. Rev. \textbf{D54}:6782, 1996.
\newblock \doi{10.1103/PhysRevD.54.6782}.
\newblock \eprint{hep-ph/9607447}.

\bibitem{Barbieri:1992dq}
R.~Barbieri, M.~Beccaria, P.~Ciafaloni, G.~Curci and A.~Vicere.
\newblock \enquote{{Two loop heavy top effects in the Standard Model}}.
\newblock Nucl. Phys. \textbf{B409}:105, 1993.
\newblock \doi{10.1016/0550-3213(93)90448-X}.

\bibitem{Barbieri:1992nz}
R.~Barbieri, M.~Beccaria, P.~Ciafaloni, G.~Curci and A.~Vicere.
\newblock \enquote{{Radiative correction effects of a very heavy top}}.
\newblock Phys. Lett. \textbf{B288}:95, 1992.
\newblock \doi{10.1016/0370-2693(92)91960-H}.
\newblock \eprint{hep-ph/9205238}.

\bibitem{Fleischer:1994cb}
J.~Fleischer, O.~V. Tarasov and F.~Jegerlehner.
\newblock \enquote{{Two loop large top mass corrections to electroweak
  parameters: Analytic results valid for arbitrary Higgs mass}}.
\newblock Phys. Rev. \textbf{D51}:3820, 1995.
\newblock \doi{10.1103/PhysRevD.51.3820}.

\end{thebibliography}
  \bibliographystyle{./bibstyles/kpinunu_Xt}


\end{document}